\documentclass[aps,prl,showpacs,twocolumn,floats,superscriptaddress,floatfix]{revtex4}
\usepackage{amsfonts,amssymb,amsmath}
\usepackage{graphics,epsfig}
\begin{document}
\title{Is Multiscaling an Artifact in the Stochastically Forced Burgers
Equation?}
\author{Dhrubaditya Mitra} 
\affiliation{Centre for Condensed Matter Theory, 
Department of Physics, Indian Institute of Science, 
Bangalore 560012, India}
\affiliation{ D\'epartement Cassiop\'ee,
 Observatoire de la C\^{o}te d'Azur, BP4229, 06304 Nice Cedex 4,
 France }
\author{J\'er\'emie Bec}
\affiliation{ D\'epartement Cassiop\'ee,
 Observatoire de la C\^{o}te d'Azur, BP4229, 06304 Nice Cedex 4,
 France }
\affiliation{Dipartimento di Fisica, Universit\`{a} La Sapienza, 
             P.zzle Aldo Moro 2, 00185 Roma, Italy }
\author{ Rahul Pandit}
\altaffiliation[Also at ]{Jawaharlal Nehru Centre For Advanced
Scientific Research, Jakkur, Bangalore, India}
\affiliation{Centre for Condensed Matter Theory, 
Department of Physics, Indian Institute of Science, 
Bangalore 560012, India}
\author{Uriel Frisch}
\affiliation{ D\'epartement Cassiop\'ee,
 Observatoire de la C\^{o}te d'Azur, BP4229, 06304 Nice Cedex 4,
 France }
\date{\today}
\begin{abstract}
We study turbulence in the one-dimensional Burgers equation 
with a white-in-time, Gaussian random force that has a 
Fourier-space spectrum $\sim 1/k$, where $k$ is the wave number. From 
very-high-resolution numerical simulations, in the limit of vanishing 
viscosity, we find evidence for multiscaling of velocity structure 
functions which cannot be falsified by standard tests. We find a 
new artifact in which logarithmic corrections can appear disguised 
as anomalous scaling and conclude that bifractal scaling is likely.
\end{abstract}
\pacs{47.27 Gs, 05.45-a, 05.40-a}
\maketitle
Homogeneous, isotropic fluid turbulence is often characterized by the 
order-$p$ velocity structure functions
$S_p(\ell) = \langle [\{{\vec v}({\vec x}+{\vec \ell}) -
{\vec v}({\vec x})\}\cdot(\frac{{\vec \ell}}{\ell})]^p \rangle$,
where ${\vec v}({\vec x})$ is the velocity at the point ${\vec x}$
and the angular brackets denote an average over the statistical
steady state of the turbulent fluid. For separations
$\ell$ in the inertial range, $\eta_d \ll \ell \ll L$,
one has $S_p(\ell) \sim \ell^{\zeta_p}$. Here $\eta_d$ is the small 
length scale at which dissipation becomes important; $L$ is the large 
length scale at which energy is fed into the fluid. The 1941 theory (K41) 
of Kolmogorov~\cite{kol41} predicts {\it simple scaling} with exponents
$\zeta^{K41}_p =p/3$. By contrast, experiments and direct numerical
simulations (DNS) suggest {\it multiscaling} with
$\zeta_p$ a nonlinear, monotonically increasing, convex function of 
$p$, not predictable by dimensional analysis~\cite{fri96}.  However, the Reynolds numbers achieved in DNS are 
limited, so the exponents $\zeta_p$ have to be extracted from numerical 
fits over inertial ranges that extend, at best, over a decade in $\ell$.
The processing of experimental data -- although they can
achieve much higher Reynolds numbers -- involves  other well-known 
difficulties~\cite{ans84}. It is important therefore to establish, 
or rule out, multiscaling of structure functions in simpler forms 
of turbulence, such as passive-scalar, passive-vector or 
Burgers turbulence. Significant progress, both analytical and 
numerical, has been made in confirming multiscaling in 
passive-scalar and passive-vector problems~(see, e.g., Ref.~\cite{rmp} 
for a review). The linearity of the passive-scalar and passive-vector 
equations is a crucial ingredient of these studies, so it is not clear 
how they can be generalized to fluid turbulence and the Navier--Stokes 
equation.

Here we revisit the one-dimensional, Burgers equation with 
stochastic self-similar forcing, studied earlier in 
Refs.~\cite{che95,hay97}. 
It is by far the simplest {\it nonlinear} partial differential 
equation (PDE) that has the potential to display multiscaling of 
velocity structure functions~\cite{hay97}; and it is akin to the 
Navier--Stokes equation. In particular, we investigate the statistical 
properties of the solutions to
\begin{equation}
\partial_t u + u \partial_x u = \nu \partial_{xx}u + f(x,t) ,
\label{eq:burg}
\end{equation}
in the limit of vanishing viscosity $\nu\to0$. Here $u$ is the velocity, 
and $f(x,t)$ is a zero-mean, space-periodic Gaussian random force with
\begin{eqnarray}
\langle{\hat f}(k_1,t_1){\hat f}(k_2,t_2)\rangle =
        2D_0|k|^{\beta}\delta(t_1-t_2)\delta(k_1+k_2) 
\label{eq:force}
\end{eqnarray}
and ${\hat f}(k,t)$ the spatial Fourier transform of $f(x,t)$. We 
restrict ourselves to the case $\beta = -1$ and assume spatial 
periodicity of period $L$.  Earlier studies~\cite{che95,hay97} suggested 
that Eqs. (\ref{eq:burg}) and (\ref{eq:force}), with $\beta = -1$, show a 
nonequilibrium statistical steady state with {\it bifractal scaling}:  
this means that velocity structure functions of order $p \leq 3$ exhibit 
self-similar scaling with exponents $p/3$ and implies a K41-type $-5/3$ energy 
spectrum, predictable by dimensional analysis, whereas
those of order $p \ge 3$ have exponents all equal to unity
being dominated by the finite number of shocks, with
$O(L^{1/3})$ strength, typically present in the periodic domain; 
this bifractal scaling is somewhat similar to that observed when the 
Burgers equation is forced only at large spatial scales~\cite{ekms97,fri00}. 
\begin{figure*}
\begin{minipage}[t]{0.31\linewidth}
\includegraphics[width=\linewidth]{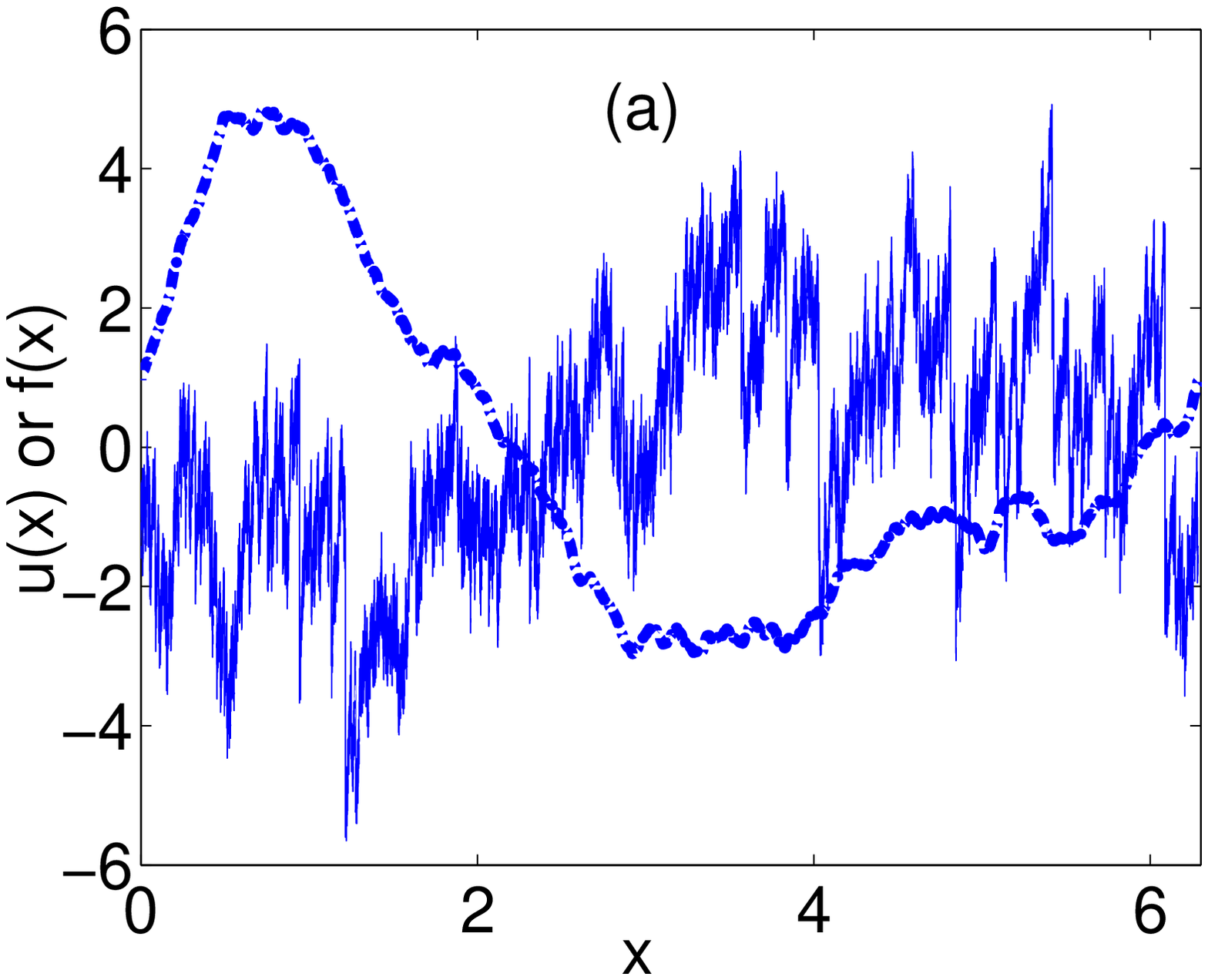}
\end{minipage} \hfill
\begin{minipage}[t]{0.31\linewidth}
\includegraphics[width=\linewidth]{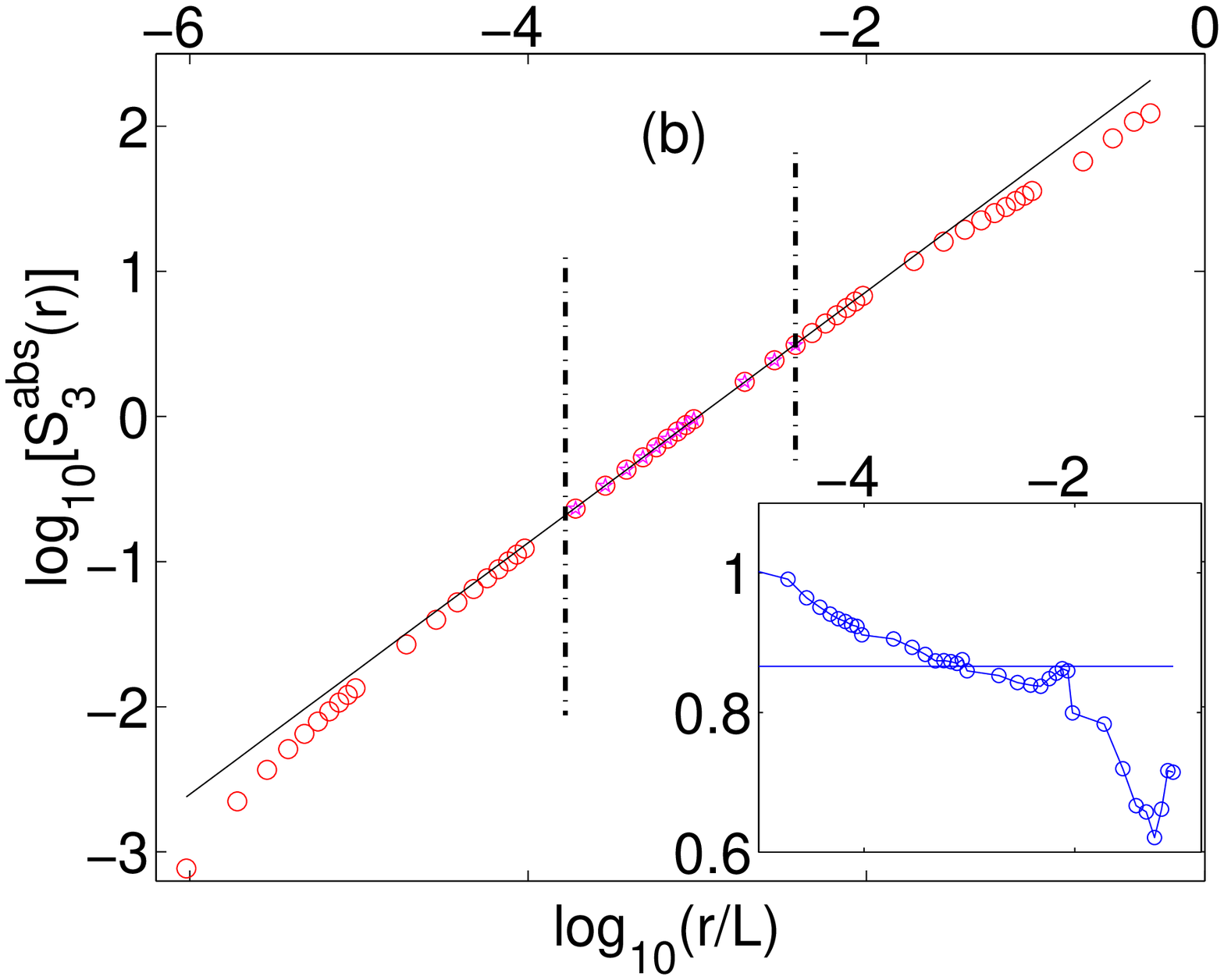}
\end{minipage} \hfill
\begin{minipage}[t]{0.31\linewidth}
\includegraphics[width=\linewidth]{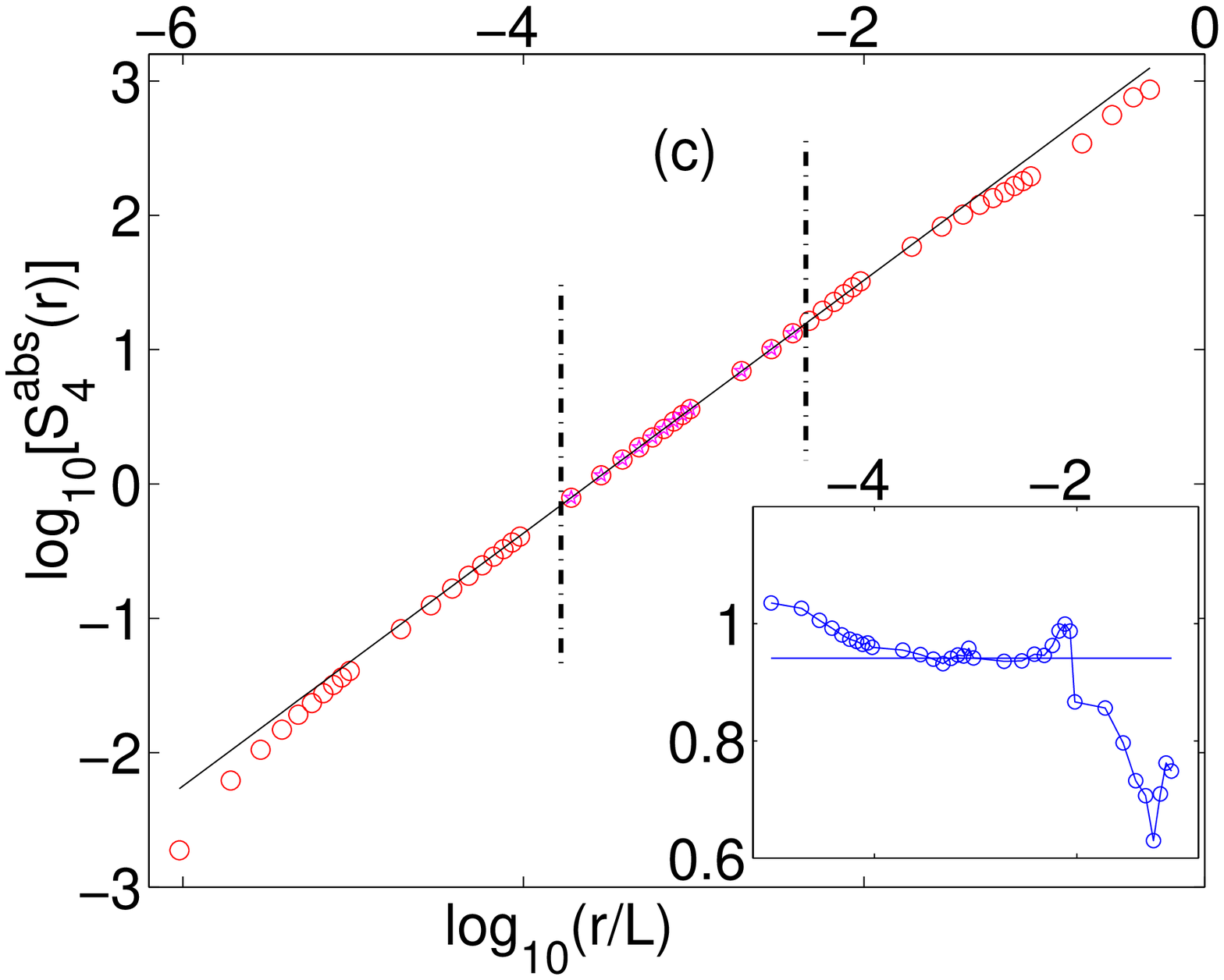}
\end{minipage}
\caption{\small
(a) Representative snapshots of the force $f$ and the velocity $u$ 
(jagged line), in the statistically stationary r\'egime; 
the velocity develops small-scale fluctuations much stronger than those
present in the force.
Log-log plots of the structure function $S^{\rm abs}_p(r)$ versus $r$ for $N = 2^{20}$ and (b) $p = 3$ and (c) $p =
4$. The straight line indicates the least-squares fit to the range of 
scales limited by the two vertical dashed lines in the plots. The resulting 
multiscaling exponents $\xi_p$ (see text) are shown by horizontal lines in the 
insets with plots of the local slopes versus $r$.} 
\label{fig:first}
\end{figure*}

\begin{figure*}
\begin{minipage}[t]{0.31\linewidth}
\includegraphics[width=\linewidth]{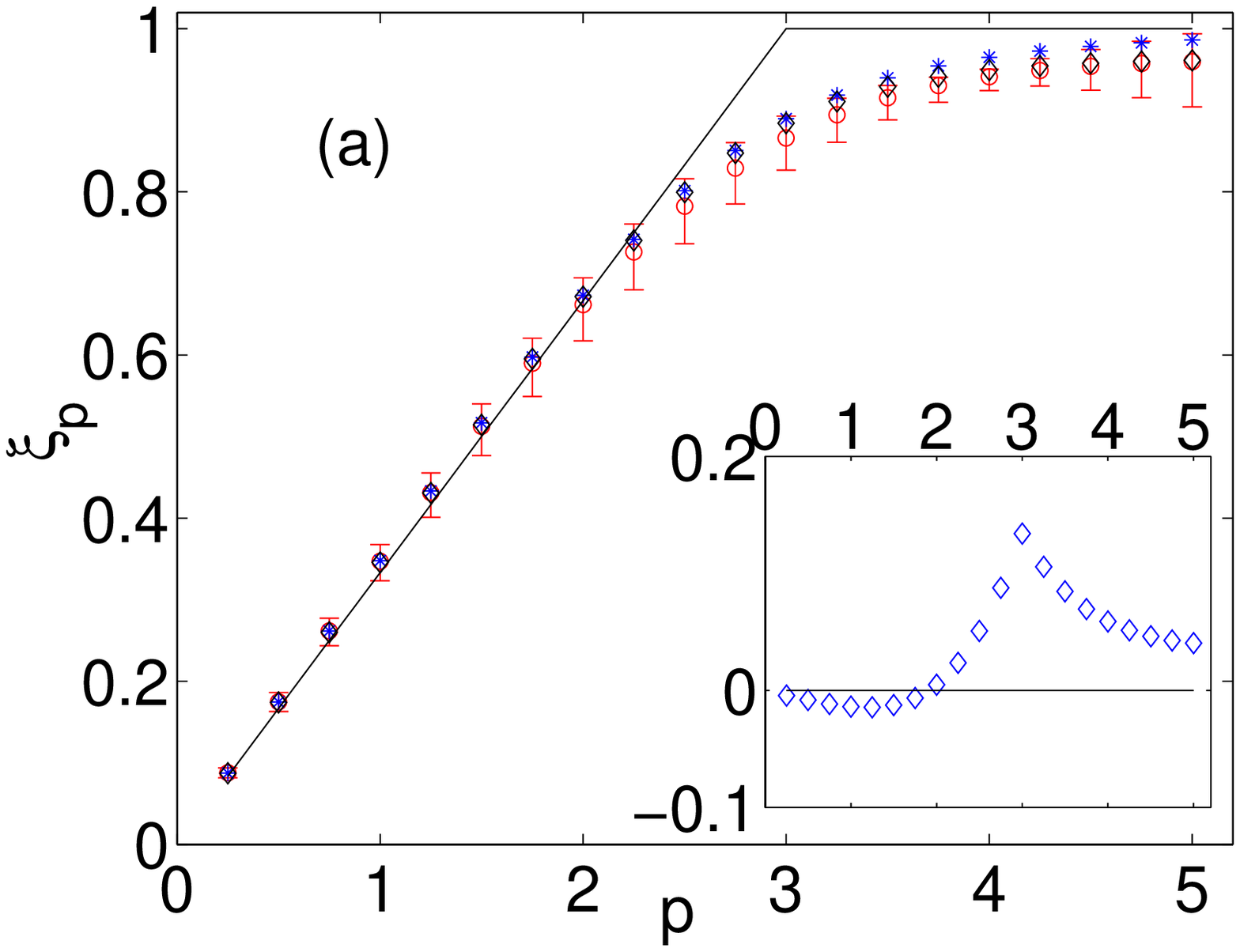}
\end{minipage} \hfill
\begin{minipage}[t]{0.31\linewidth}
\includegraphics[width=\linewidth]{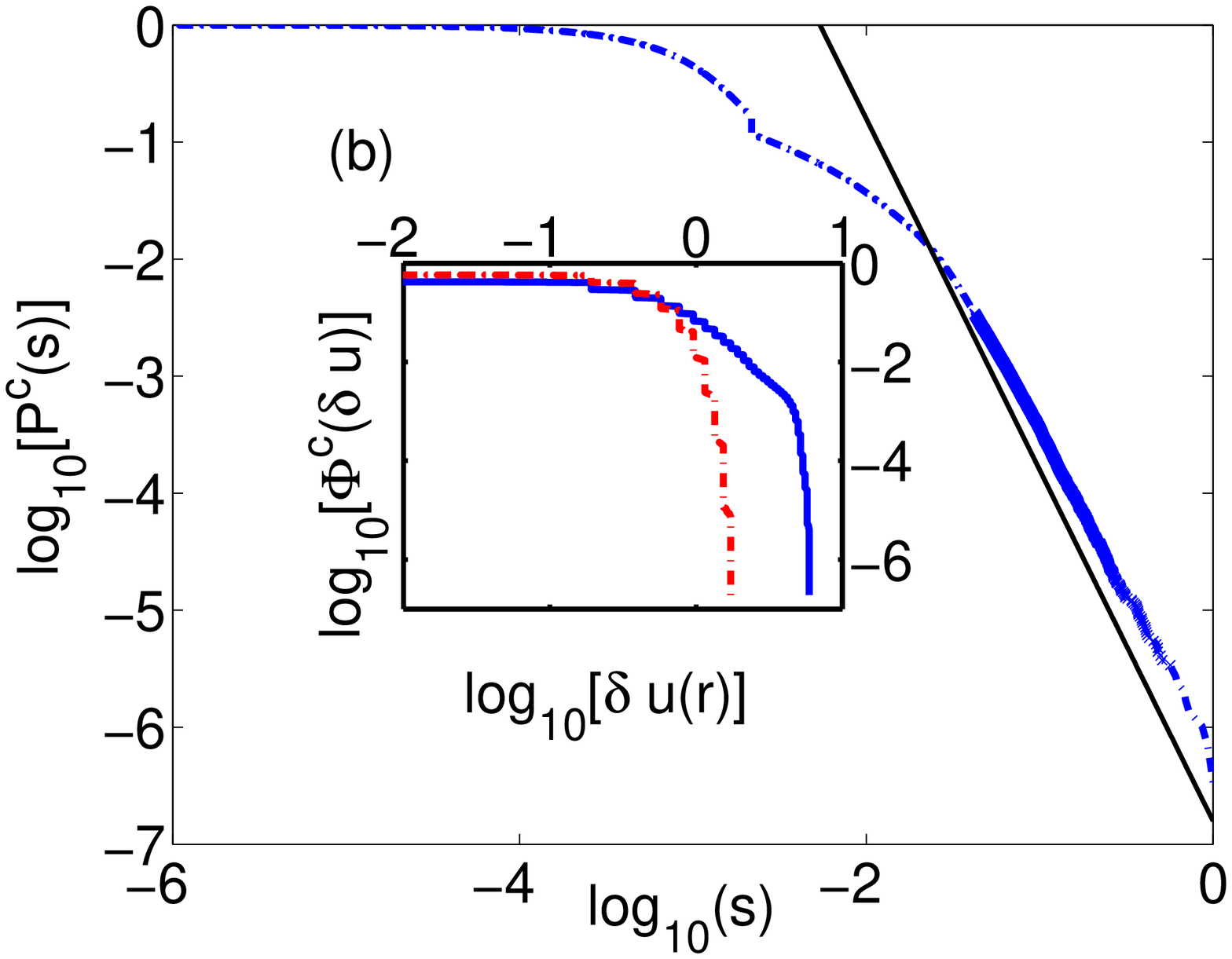}
\end{minipage}
\begin{minipage}[t]{0.31\linewidth}
\includegraphics[width=\linewidth]{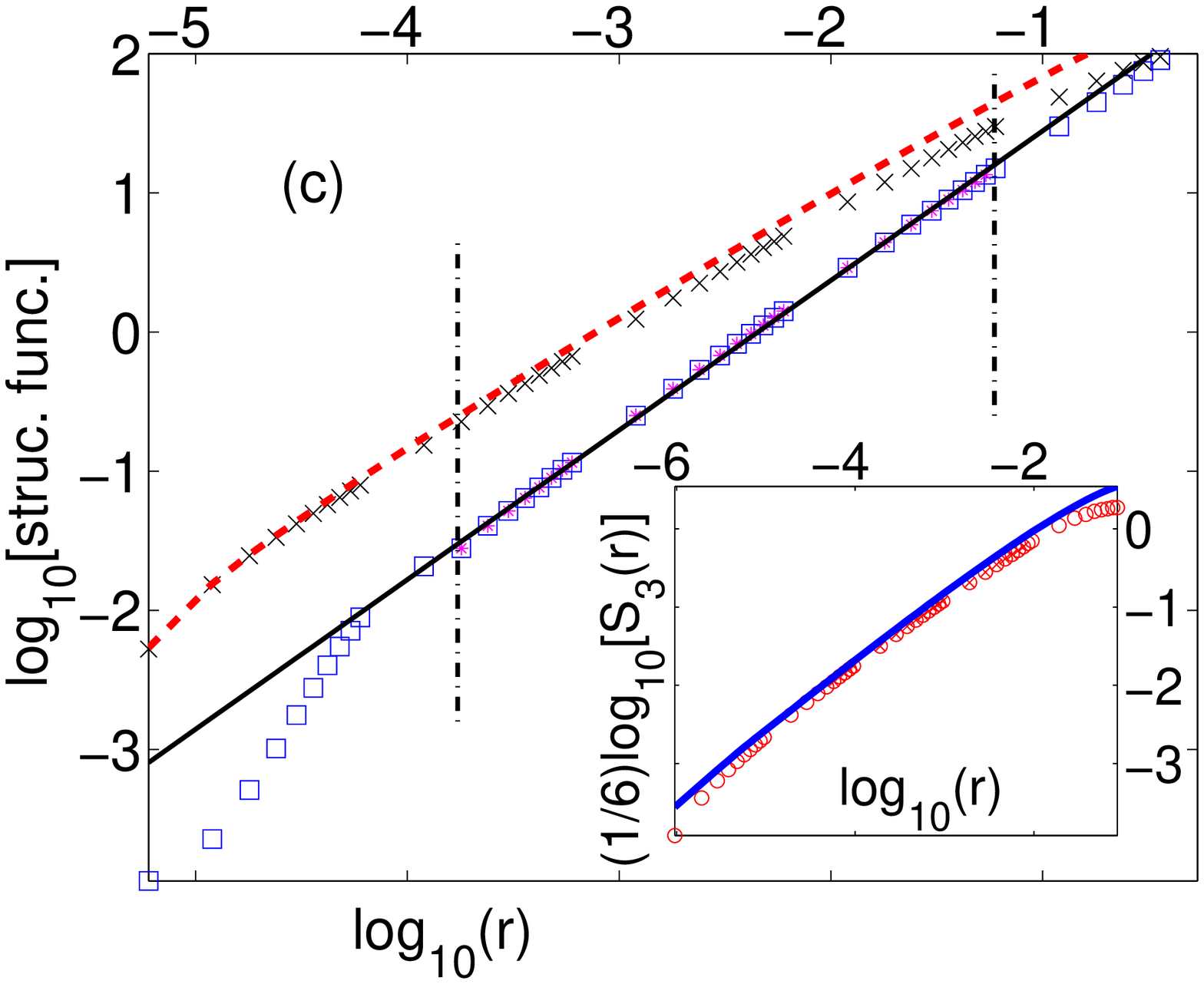}
\end{minipage} \hfill
\caption{\small
(a) The multiscaling exponents $\xi_p$ versus order $p$ for 
Eqs. (\ref{eq:burg}) and (\ref{eq:force}) with 
$N = 2^{16} (\diamond), \,2^{18} (\ast)$, and $2^{20} (\circ)$ 
grid points. Error bars (see text) are shown for the case 
$N = 2^{20}$. The deviation of $\xi_p$ from the exponents for 
bifractal scaling (full lines), shown as an inset, suggest naive 
multiscaling.
(b) Log-log plots of the cumulative probability distribution function 
$ P^c(s)$ versus shock strengths $s$ obtained from an average over
1000 snapshots. A least-squares fit to the form 
$ P^c(s) \sim s^{\gamma}$, for the dark points in the range 
$-5 \lesssim \log_{10}[P(s)] \lesssim -2.5$, yields 
$\gamma = -2.70 $; the simple-scaling prediction $\gamma = -3$ is 
indicated by the straight line.  The inset shows log-log plots of the 
cumulative probability distribution function, 
$\Phi^c[\delta u(r)]$, (dashed line : positive $\delta u$,
continuous line : negative $\delta u$) versus the velocity difference 
$\delta u(r)$ for length scale $r=800 \delta x$.   
(c) Log-log plots of $S_3(r)$ (crosses), $S_3^{\rm abs}(r)$ (dashed line)
and $\langle (\delta^{+}u)^3\rangle $ (squares) versus $r$. 
The continuous line is a least-square fit to the range of points
limited by two vertical dashed lines in the plot.  
Inset: An explicit check of Eq.~(\ref{eq:exact}) from our 
simulations, plotted on a log-log scale. The dashed line is the 
right-hand side of Eq.~(\ref{eq:exact}); the left-hand side of this 
equation has been obtained for $N = 2^{20}\;(\circ)$ (run B1).}
\label{fig:second}
\end{figure*}

We overcome the limitations of these earlier studies~\cite{che95,hay97} by 
adapting the algorithm of Refs.~\cite{nou94,bec00} to develop a 
state-of-the-art technique for the numerical solution of Eqs. (\ref{eq:burg}) 
and (\ref{eq:force}), in the $\nu \to 0$ limit.  This yields velocity profiles 
(Fig.~\ref{fig:first}~a) with shocks at all length scales resolved. 
Structure functions [Figs.~(\ref{fig:first}~b) and (\ref{fig:first}~c)] 
exhibit power-law behavior over nearly three decades of $r$; this is more 
than two decades better than in Ref.~\cite{che95}. 
In principle it should then be possible to measure the scaling
exponents [Figs.~(\ref{fig:first}~b)] with enough accuracy to decide
between bifractality and multiscaling. A naive analysis
[Fig.~(\ref{fig:second}~a)] does suggest multiscaling
\footnote{In Ref.~\cite{hay97} the authors argued that
Eqs.~(\ref{eq:burg}) and (\ref{eq:force}), with $-1 < \beta < 0$,
could lead to {\it genuine multifractality}.  Their results, obtained
by a pseudospectral DNS with a finite viscosity, had a scaling range
far smaller than can be achieved now. Our results for values of
$\beta$ in this range will be reported elsewhere.}.  However, given
that simple scaling or bifractal scaling can sometimes be mistaken for
multiscaling in a variety of models~\cite{kun99,aur97}, it behooves us
to check if this is the case here.  We describe below our numerical
procedure and the various tests we have carried out. 

In our simulations we use $L=2\pi$ and $D_0=1$ without loss of generality. 
The spatial mesh size is $\delta x = L/N$ where $N$ is the number of grid points.
To approximate the forcing, we use the ``kicking'' strategy 
of Ref.~\cite{bec00} in which the white-in-time force is approached
by shot noise.  Between successive kicks we evolve the velocity 
by using the following  
well-known result on the solutions to the unforced Burgers equation 
in the limit of vanishing viscosity (see, e.g., Ref.~\cite{fri00}): 
the velocity potential $\psi$ (such that $u = -\partial_x \psi$) 
obeys the maximum principle
\begin{equation}
\psi(x,t^{\prime}) = \max_y \left[\psi(y,t) -
\frac{(x-y)^2}{2\, (t^{\prime} -t)} \right]; \quad t^{\prime} > t.
\label{eq:psi}
\end{equation}
The search for the maxima in Eq.~(\ref{eq:psi}) requires 
only $O(N\log_2 N)$ operations~\cite{nou94} because, under 
Burgers dynamics, colliding Lagrangian particles form 
shocks and do not cross each other. At small scales we 
want to have unforced Burgers dynamics with well-identifiable
shocks. At least four mesh points are needed for unambiguous
identification of a shock; since  the maximum wavenumber is $N/2$, we
set ${\hat f}(k,t) = 0$ beyond an ultra-violet cutoff $\Lambda = N/8$. 

Specifically, at time $t_n=n\delta t$ we add $f_n(x)\sqrt{\delta t}$
to the Burgers velocity $u(x,t)$, where the $f_n(x)$s are independent
Gaussian random functions with zero mean and a Fourier-space spectrum
$\sim 1/k$, for $k < \Lambda$.  
The time step $\delta t$ is chosen to satisfy 
the following conditions $ (\delta
x/2u_0) < \delta t \simeq (1/L\Lambda)^{2/3}(L/u_0)$, where $u_o$ is
the characteristic velocity difference at large length
scales~$O(L)$. The first  ensures that a typical Lagrangian
particle moves at least half the mesh-size in time $\delta t$
(otherwise, it would stay put)\footnote{This condition is opposite
to the Courant--Friedrich--Lewy-type condition, which is not needed with our 
algorithm}; the second, which expresses that $\delta t$ is comparable to the 
turnover times at the scale $L/\Lambda$, guarantees that,  at scales larger 
than $L/\Lambda$ the time-step, $\delta t$, is small compared to all 
dynamically significant times, but still permits the formation of individual 
shocks at smaller scales. Finally, as our simulations are very long, for the 
stochastic force we use a good-quality random-number generator with a long 
repeat period of $2^{70}$ due to Knuth~\cite{knu99}. The main characterisitics 
of the runs performed are summarized in Table~\ref{table:para}.
\begin{table}
\framebox{\begin{tabular}{c|c|c|c|c|c|c}
Run & $N$ & $\delta t$ & $\Lambda$ & $\tau_L$ & $T_{tr}$ & $T_{av}$ \\
\hline
B1 & $2^{20}$ & $5 \times 10^{-4}$ & $2^{17} $ & $1.0 $ & $ 2.0 $  & $22$   \\
B2 & $2^{18}$ & $1 \times 10^{-4}$ & $2^{15} $ & $1.0 $ & $ 2.0 $  & $20$   \\
B3 & $2^{16}$ & $1 \times 10^{-4}$ & $2^{13} $ & $1.0 $ & $ 2.0 $  & $120$   \\

\end{tabular}}
\caption{ Different parameters used in our runs B1, B2 and B3. 
$\tau_L\equiv L/u_0$ is the equivalent of the large-eddy-turnover time. 
Data from $T_{tr}$ time steps are discarded so that transients die down. 
We then average our data over a time $T_{av}$.}
\label{table:para}
\end{table}
 
In addition to the usual structure functions, we have used
$S_p^{\rm abs}(r)$, defined by 
\begin{eqnarray}
S^{\rm abs}_p(r) &\equiv& \langle |\delta u(x,r)|^p \rangle 
\sim r^{\xi_p}, \\
\delta u(x,r) &\equiv&  u(x+r) - u(x),
\label{eq:stfun} 
\end{eqnarray}  
 from which we extract the exponents $\xi_p$. 
For each value of $N$ we have calculated $\xi_p$ for $p = m/4$, with 
integers $1 \leq m \leq 20$. Figure (\ref{fig:second}~a) summarizes 
the results of our calculations concerning the  exponents
$\xi_p$, for $N = 2^{16},\;2^{18}$, and $2^{20}$; any systematic change 
in the values of these exponents with $N$ is much less than the error bars
determined by the procedure described below. 
Thus in all other plots we present data from our simulations with 
$N=2^{20}$ grid points.  The representative log-log plots of 
Figs.~(\ref{fig:first}~b) and (\ref{fig:first}~c) of $S^{\rm abs}_p(r)$ 
for $p=3$ and $4$ show power-law r\'egimes that extend over nearly three 
decades of $r/L$. We obtain our estimates for the exponents $\xi_p$ as 
follows: for a given value of $p$ we first determine the local slopes of the 
plot of $\log S^{\rm abs}_p$ versus $\log r$ by least-squares fits to all 
triplets of consecutive points deep inside the power-law r\'egime 
\footnote{Deviations from this power-law r\'egime occur at small 
values of $r$ because of the ultraviolet cutoff $\Lambda$ for the stochastic 
force~(\ref{eq:force}).}. These regions extend over one and half decade of
$r/L$ as shown in Figs.~(\ref{fig:first}~b) and (\ref{fig:first}~c).
The value of $\xi_p$ we quote [Fig.~(\ref{fig:second}~a)] is the mean of these 
local slopes; and the error bars shown are the maximum and minimum local slopes 
in these regions.  

Figure (\ref{fig:second}~a) shows that our results for $\xi_p$, indicated
by circles for $N = 2^{20}$, deviate significantly from the bifractal-scaling
prediction (full lines). As we shall see this deviation need
not necessarily imply ``multiscaling'' for the structure functions. 
We have also considered the possibility of already well-understood artifacts,
such as the role of temporal transients~\cite{kun99} and finite-size
effects which can round sharp bifractal transitions~\cite{aur97}, 
and decided that they do not play any major role in the present 
problem~\footnote{Details on such questions can
be found in the earlier version of the present paper, available at {\tt
http://xxx.lanl.gov/abs/nlin.CD/0406049v1}}. 

Consider $P^c(s)$, the cumulative probability distribution function of shock strengths 
$s$. Simple scaling arguments predict $P^c(s) \sim s^{\gamma}$, with 
$\gamma = -3$, which follows by demanding that $P^c(s)$ remain invariant 
if lengths are scaled by a factor $\lambda$ and velocities by 
$\lambda^{-1/3}$. One of the signatures of multiscaling would be 
deviations of $\gamma$ from this scaling
value. However, the following argument favors $\gamma = -3$ :
the total input energy,  
$E_{\rm in} = \int_{k_0}^{\Lambda} D(k) dk \sim \ln \Lambda$,
where $\Lambda$ is the ultra-violet cutoff. 
In the limit of vanishing viscosity, the energy 
dissipation in the Burgers equation occurs only at the shocks 
and is proportional to the cube of the shock strength~\cite{fri00},
so the total energy dissipation is 
$\Omega \sim \int_{s_{\rm min}}^{s_{\rm max}} P(s) s^3 ds$.
Here $P(s) = (dP^c(s)/ds) \sim s^{\gamma -1}$ 
is the probability distribution function~(PDF) of shock strengths 
and $s_{\rm min}$ and $s_{\rm max}$ are, respectively, the 
minimum and maximum shock strengths. 
A steady state can occur only if  $E_{\rm in}$ and $\Omega$ have
the same asymptotic properties as $\Lambda \to \infty$, i.e., 
$\Omega \sim \ln(\Lambda)$; this requires $\gamma = -3$. 
By contrast we find 
$\gamma \simeq -2.7$ [Fig.~(\ref{fig:second}~b)] by a naive least-squares 
fit to the tail of $P^c(s)$ 
\footnote{To estimate shock locations we look at groups of 
four grid points where the discretized velocity gradient changes its 
sign twice (they correspond to a ``zig-zag'' in the velocity profile).} . 
This suggests that the results of our simulation are far from the 
limit $\Lambda \to \infty$, although more than a million grid points are 
used. Hence the ``anomalous'' exponents in Fig.~(\ref{fig:second}~a) might 
well be suspect. 

To explore this further, consider the third-order structure function of 
velocity differences, without the absolute value, namely,
$S_3(r) \equiv \langle \delta u^3 \rangle$.
From Eqs. (\ref{eq:burg}) and (\ref{eq:force}) follows the
exact relation
\begin{equation}
\frac{1}{6} S_3(r) = \int_0^{r} F(y) dy,
\label{eq:exact}
\end{equation}
where $F(y)$ is the spatial part of the force correlation function, defined
by $\langle f(x+y,t^{\prime})f(x,t)\rangle = F(y)\delta(t-t^{\prime})$.
We obtain this analog of the von K\'arm\'an--Howarth relation in fluid 
turbulence by a simple generalization of the proof given in 
Ref.~\cite{bec00} for the Burgers equation forced 
deterministically at large spatial scales.  An explicit check of 
Eq.~(\ref{eq:exact}) provides a stringent test of our simulations
[Fig.~(\ref{fig:second}~c) inset]. Furthermore,  
Eq.~(\ref{eq:exact}) implies that  $S_3(r)\sim r \log (r)$ 
for small $r$ and thus should display significant curvature in a log-log 
plot, as is indeed seen in Fig.~(\ref{fig:second}~c).
By contrast $S_p^{\rm abs}(r)$ [Fig.(\ref{fig:second}~c)] displays much 
less curvature and can be fitted over nearly three decades in $(r/L)$ 
to a power law with an ``anomalous'' exponent of $0.85$.
 This anomalous behavior is probably an artifact as we now show. 
Let us define the positive (resp.,\ negative) part of the velocity
increment  $\delta^{+}u $ (resp., $\delta^{-}u $) equal
to $\delta u$ when $\delta u \ge 0$ and to zero when $\delta u < 0$
(resp., to $\delta u$ when $\delta u \le 0$ and to zero when 
$\delta u > 0$). Obviously $S_3(r) = \langle (\delta^{+}u)^3\rangle +
 \langle (\delta^{-}u)^3\rangle$, whereas $S_p^{\rm abs}(r) = 
\langle (\delta^{+}u)^3\rangle -  \langle (\delta^{-}u)^3\rangle$.
The log-log plot of 
$\langle (\delta^{+}u)^3\rangle = (1/2)[S_3(r) + S_3^{\rm abs}(r)]$ 
in Fig.~(\ref{fig:second}~c) is much straighter than those for 
$S_3(r)$ and $S_3^{\rm abs}(r)$ and leads to a scaling exponent 
$1.07 \pm 0.02$, very close to unity.
For a moment assume that $\langle (\delta^+ u)^3 \rangle $ 
indeed has a scaling exponent of unity. Given that $S_3(r)$ has, 
undoubtedly, a logarithmic correction, it follows that 
$S_3^{\rm abs}(r)$ has
(except for a change in sign) the same logarithmic correction in its 
leading term (for small $r$) but \emph{differs by a subleading correction
proportional to $r$}. This subleading correction
is equivalent to replacing $r \log (r)$ by $r \log (\lambda r)$ for
a suitably chosen factor $\lambda$. In a log-log plot this shifts
the graph away from  where it is most curved and thus makes it 
straighter, albeit with a (local) slope which is  not unity.

An independent  check of $\langle (\delta^+ u)^3 \rangle \sim r $  
is obtained by plotting the cumulative probabilities, $\Phi^c$, of
positive and negative velocity increments (for a separation 
$r=800\delta x$) in Fig.~(\ref{fig:second}~b)
\footnote{Similar plots from lower-resolution viscous spectral 
simulations were obtained in Ref.~\cite{che95}.}. 
For positive increments $\Phi^c$ falls 
off faster than any negative power of $\delta u$, but, for 
negative ones, there is a range of increments over which 
$\Phi^c \sim |\delta u|^{-3}$, the same $-3$ law seen in $P^c(s)$ 
earlier. Indeed the negative increments are dominated by the 
contribution from shocks. Just as $P^c(s)$ has cutoffs $s_{\rm min}$ 
and $s_{\rm max}$, $\Phi^c$ has cutoffs $u^-_{\rm min}(r)$ and 
$u^-_{\rm max}(r)$ for negative velocity increments.
Since $\Phi^c$ falls off as $|\delta u|^{-3}$, $u^-_{\rm max}$ can be 
taken to be $\infty$; furthermore as the PDF of velocity differences,
 $\Phi(\delta u) \equiv d(\Phi^c)/d(\delta u)$, must be normalizable, 
we find $u^-_{\rm min}(r) \sim r^{1/3}$. We now know enough about the 
form of $\Phi$ to obtain, in agreement with our arguments above, 
that $S_3(r) \approx -Ar\ln(r) + Br$ and  
$S^{\rm abs}_3(r) \approx Ar\ln(r) + Br$,
whence $\langle (\delta^+ u)^3 \rangle \approx Br$.  The presence 
of this cutoff yields a logarithmic term in both $S_3$ and 
$S_3^{\rm abs}$ but with different sign agreeing with the arguments 
given in the previous paragraph. 

By a similar approach, we find   
$S_4(r) \approx C r - D r^{4/3}$, 
where $C$ and $D$ are two positive constants.  The negative sign 
before the sub-leading term~$(r^{4/3})$ is crucial. It implies 
that,  for any finite $r$, a naive power-law fit to $S_4$ can yield
a scaling exponent less than unity. The presence of sub-leading, 
power-law terms with opposite signs also explains the small apparent 
``anomalous'' scaling behavior observed  for other values of $p$ in 
our simulations.  A similar artifact involving two competing power-laws 
has been described in Ref.~\cite{bif04}. 

In conclusion, we have performed  very-high-resolution numerical 
simulations of the stochastically forced Burgers equation with a $1/k$ 
forcing spectrum. A naive interpretation of our data shows apparent 
multiscaling phenomenon. But our detailed analysis has 
identified a hitherto-unknown numerical artifact by which simple
biscaling can masquerade as multiscaling. Our work illustrates 
that the elucidation of multiscaling in spatially extended
nonlinear systems, including the Navier--Stokes equation, requires  
considerable theoretical insight that must supplement state-of-the-art 
numerical simulations and experiments.  

We thank E. Aurell, L. Biferale, A. Lanotte, and V. Yakhot for 
discussions.  This research was supported by the Indo-French Centre 
for the Promotion of Advanced Research (Project 2404-2), by 
CSIR (India), and by the European Union under contracts 
HPRN-CT-2000-00162 and HPRN-CT-2002-00300. Additional computational 
resources were provided by CHEP (IISc).

\printfigures
\printtables
\end{document}